\documentclass[usenatbib]{mn2e}
\usepackage{graphicx}

\def\etal{{\it et\thinspace al\/}}

\def\kpch{\,{h^{-1}{\rm kpc}}}
\def\mpch{\,h^{-1}{\rm {Mpc}}}

\def\msun{M_\odot}

\def\lcdm{$\Lambda$CDM }

\bibpunct[,]{(}{)}{;}{a}{}{,}

\begin{document}

\title[On the Distribution of Haloes, Galaxies and Mass]{On the Distribution of Haloes, Galaxies and Mass}
\author[R. Casas-Miranda \etal]{R. Casas-Miranda$^{1}$,
H.J. Mo$^{1}$, Ravi K. Sheth$^{2}$  and G. B\"orner$^{1}$ \\
1. Max-Planck-Institut f\"ur Astrophysik, Karl Schwarzschildstr. 1,
85741, Garching, Germany\\
2. NASA/Fermilab Astrophysics Group, MS 209, Batavia, IL 60510-0500 \\
\smallskip
Email: casas@mpa-garching.mpg.de, hom@mpa-garching.mpg.de, 
sheth@fnal.gov, grb@mpa-garching.mpg.de}
\date{Accepted 2002 February 1; In original form 2001 May 1}
\maketitle

\begin{abstract}
The stochasticity in the distribution of dark haloes in the 
cosmic density field is reflected in the distribution 
function $P_V(N_h|\delta_m)$ which gives the probability
of finding $N_h$ haloes in a volume $V$ with mass density
contrast $\delta_m$. We study the properties of this 
function using high-resolution $N$-body simulations, and 
find that $P_V(N_n|\delta_m)$ is significantly non-Poisson.
The ratio between the variance and the mean goes from 
$\sim 1$ (Poisson) at $1+\delta_m\ll 1$ to $<1$ (sub-Poisson) at 
$1+\delta_m\sim 1$ to $>1$ (super-Poisson) at 
$1+\delta_m\gg 1$. The mean bias relation is found to be well
described by halo bias models based on the Press-Schechter
formalism. The sub-Poisson variance can be explained as a result of
halo-exclusion while the super-Poisson variance at high $\delta_m$ may
be explained as a result of halo clustering. A simple phenomenological
model is proposed to describe the behavior of the variance as a
function of $\delta_m$. Galaxy distribution in the cosmic density
field  predicted by semi-analytic models of galaxy  formation shows
similar stochastic behavior. We discuss the implications of the
stochasticity in halo bias to the modelling of higher-order moments of
dark haloes and of galaxies.
\end{abstract}  


\section {Introduction}  

In the current scenario of galaxy formation, galaxies are 
assumed to form by the cooling and condensation of gas within
dark matter haloes \citep[e.g.][]{WhiteR:78, WhiteF:91}. The problem of
galaxy clustering in space can therefore be approached by 
understanding the spatial distribution of dark haloes and the 
formation of galaxies in individual dark haloes. 
This approach to the problem of galaxy spatial clustering
is very useful because the formation and clustering properties 
of dark haloes can be modelled relatively reliably due to
the simplicity of the physics involved (gravity only) 
and because realistic models of galaxy formation in dark haloes can now be
constructed using semi-analytic models \citep[e.g.][]{Kauffmann:99,
Cole:00, SomervilleP:99}. Indeed, there are quite a few 
recent investigations attempting to model galaxy clustering 
based on the halo scenario
\citep[e.g.][]{JingMB:98, MaF:00, Scoccimarro:01, PeacockS:00,
Seljak:00, ShethMT:01}.

Based on the Press-Schechter formalism \citep[][]{PressS:74}
and its extensions \citep[][]{LaceyC:94}, \citet[][]{MoW:96} (hereafter
MW) developed a model for the mean bias relation for 
dark haloes. Their model and its extension 
based on ellipsoidal collapse \citep[][]{ShethMT:01} have been
extensively tested by N-body simulations \citep[e.g. MW;][]{MoJW:96,
JingMB:98, ShethT:99, Governato:99, Colberg:00}. High-order  
moments of the halo distribution have also been modeled by 
\citet[][]{MoJW:97} based on a deterministic bias relation. These
authors showed that the model works on large scales
in comparison with N-body simulations.  
Nevertheless, the effect of stochasticity may be
important in these high-order statistics as well as in the full 
distribution function of haloes. In fact, the non-Poissonian behavior of 
the bias relation is already emphasized in the original paper of
MW; in particular, MW pointed out that halo-exclusion can cause 
sub-Poisson variance. \citet[][]{ShethL:99} showed how the effects of 
stochasticity could be incorporated, easily and efficiently, into the 
analysis of the higher order moments.

Recently \citet[][]{Somerville:01} used $N$-body simulations to
study the stochasticity and non-linearity of the bias relation 
based on the formalism developed by \citet[][]{DekelL:99}. 
They analyzed the bias relation for haloes with masses larger than 
$1.0\times 10^{12}~h^{-1}~\msun$ in
spherical volumes of radius $8\mpch$. Our present work is quite 
closely related to theirs but contains several distinct aspects.
First of all, our analysis is focused on the distribution 
function $P_V(N_h\vert\delta_m)$, which gives the probability
of finding $N_h$ haloes in a volume $V$ with mass density
contrast $\delta_m$ [$\delta_m \equiv {\rho\over {\overline{\rho}}}
-1$, where $\rho$ is the mass density and ${\overline{\rho}}$ is the
mean mass density]. As we will show later, this function completely 
specifies the relation between the spatial distribution of 
haloes and that of the mass in a statistical sense. Second, 
our analysis covers a wider range of halo masses and a larger 
range of volumes for the counts-in-cells.
Finally, we attempt to develop a theoretical model to describe
the stochasticity of the bias relation. This theoretical model
is based on the mean bias relation given in MW  and on the variance
model given in \citet[][]{ShethL:99}. As we will see below, the Sheth
\& Lemson model fails in high mass density regions, where
gravitational clustering becomes important. One of the main purposes
of this paper is to show that a simple modification of the Sheth \&
Lemson formulae for the variance allows one to make accurate
predictions even in dense regions. \citet[][]{TaruyaS:00} have
proposed a model for the stochasticity in halo bias relation based on
the formation-epoch distribution of dark haloes, an approach very
different from ours. 

The paper is organized as follows. In Section \ref{sec:bias} we
introduce the bias relation based on the conditional probability and
present a phenomenological model to describe the behavior of the
variance as a function of the local density contrast. In Section
\ref{sec:testing} we present the numerical data used and study the
mean and variance of the bias relation. We discuss and summarize 
our results in Section \ref{sec:summary}.

\section{The Halo-Mass Bias Relation} 
\label{sec:bias}

\subsection{The Concept of Bias}
In the following we shall introduce some important concepts we will use
throughout the paper. For simplicity we introduce them for the
case of dark matter haloes, without loss of generality.

Let us define $\rho$ as the mass density smoothed in regions of some
given volume $V$. The mass density contrast ($\delta_m$) in this volume
is then defined as:

\begin{equation}
\delta_m \equiv {\rho\over {\overline{\rho}}}-1,
\label{eq:moverd}
\end{equation}

\noindent where ${\overline\rho}$ is the mean mass density in the
universe. In the same way, if $N_h$ and $\overline{N}_h$ correspond to the number of
dark matter haloes and  to the mean number of dark matter haloes in
the volume $V$, respectively, the number density contrast of dark
matter haloes ($\delta_h$) is given by
\begin{equation}
\delta_h \equiv { N_h \over {\overline{N}_h}}-1.
\label{eq:hoverd}
\end{equation}

The relationship between $\delta_h$ and $\delta_m$ is widely known as
the halo-mass bias. A general way to represent this bias
relation is to express the halo number density contrast as
a function of the mass density contrast 

\begin{equation}
\delta_h(V) \equiv F(\delta_m(V)).
\label{eq:bias_general}
\end{equation}
Since the mass and halo number density contrasts are defined in a
given volume, the bias relation defined in equation \ref{eq:bias_general} is
assumed to be a function of the local fields and thus called ``local
halo-mass bias''. The exact form of the function $F(\delta_m(V))$
depends on how the objects in consideration form in the cosmic density
field.

The bias function $F(\delta_m(V))$ can be of several kinds. If the
bias function is a linear function of the local mass density contrast
$\delta_m(V)$ then it is said that one has linear bias. On the other
hand, if the bias function is a non-linear function of the local mass
density contrast $\delta_m(V)$ then one has non-linear bias. A special
case occurs when $F(\delta_m(V))=\delta_m(V)$, in this case the local
halo field is unbiased respect to the local mass density field.  

The above mentioned biasing schemes correspond to the case of 
deterministic bias. A more general description of the
biasing process corresponds to a stochastic one, i.e, there is a
non-zero dispersion of the bias around its mean and the bias
relation is described in a probabilistic way. There are two factors
which can cause such dispersion. First, for a given region, not only
$\delta_m$, but also other local properties, such as clumpiness and
deviation from spherical symmetry, can affect the halo number in the
region. Second, some non-local effects, such as the large-scale tidal
field, can affect the formation of haloes. In general the
stochasticity of the bias relation can be described by the conditional
distribution function, $P_V(N_h \vert\delta_m)$, which gives the
probability of finding $N_h$ haloes in a volume $V$ with mass density
contrast $\delta_m$. The deterministic bias scheme is recovered when
the conditional probability can be well approximated by a delta
function.

\subsection {The conditional probability} 

Dark matter haloes are formed in the cosmological density field
due to nonlinear gravitational collapse. In general, the halo
density field is expected to be correlated with the underlying 
mass density field. Thus, if we denote by $\delta_m$ the matter 
density fluctuations field and by $N_h$ the field of halo number 
(where both fields are smoothed in regions of some given volume),
$N_h$ and $\delta_m$ are related. We refer to this relation as the halo 
bias relation (see last subsection). Since in general the halo number
in a volume depends not only on the mean mass density but also on
other properties of the volume, the relation between $N_h$ and
$\delta_m$ is not expected to be deterministic. It must be
stochastic. The stochasticity of the bias relation can be described by
the conditional distribution function, $P_V(N_h \vert\delta_m)$, which
gives the probability of finding $N_h$ haloes in a volume $V$ with mass
density contrast $\delta_m$. 
This conditional probability completely specifies the 
relation between the mass and halo density fields in a statistical
sense. Indeed, once $P_V(N_h\vert\delta_m)$ is known, the full 
count-in-cell function $P_V(N_h)$ for haloes can be obtained 
from the mass distribution function $P_V(\delta_m)$ through 
\begin{equation}
P_V(N_h) = \int_{-\infty}^{\infty}{P_V(N_h \mid \delta_m)
P_V(\delta_m)\,d\delta_m}.
\label{eq:pdf}
\end{equation}

The form of $P_V(N_h\vert\delta_m)$ depends on how 
dark haloes form in the cosmological density field
and is not known {\it a priori}. The simplest stochastic  
assumption is that it is Poissonian. This assumption 
is in fact used in almost all interpretations 
of the moments of galaxy counts in cells \citep[c.f.][]{Peebles:80}, where 
terms of Poisson shot noise are subtracted to obtain
the correlation strength of the underlying density field. 
However, this assumption is not solidly based,
and so it is important to examine if other assumptions 
on the form of $P_V(N_h\vert\delta_m)$ actually work better
for dark haloes. In this paper, we test other three models of 
$P_V(N_h\vert\delta_m)$, along with the Poisson model.
These are the Gaussian model, the Lognormal model, and 
the thermodynamical model. The last model was developed by 
\citet[][]{SaslawH:84} for the distribution of galaxies. 

\subsection{A Model for the Halo-Mass Bias Relation} 
\label{sec:model}  

To second order, the probability distribution 
function $P_V(N_h\vert\delta_m)$ is described by the mean 
bias relation $N=N(\delta_m)$ and the variance
$\sigma^2 \equiv \langle N^2\vert\delta_m\rangle$.
MW developed a model for the mean bias relation of haloes based on the
spherical collapse model. Their model works well for massive
haloes and an extension of it by \citet[][]{ShethMT:01} based on
ellipsoidal collapse may work better for low mass haloes.
 
\citet[][]{ShethL:99} have presented a model for the variance of the bias
relation which accounts for the halo exclusion due to the finite size
of haloes (i.e. two different haloes can not occupy the same
volume). They showed that their model was able to 
describe the first and second moments of the halo distribution from
scale-free N-body simulations. Nevertheless the model is expected to
fail when the underlying clustering makes a significant 
contribution to the variance. As an amendment, 
we introduce an additional term accounting for the clustering 
of haloes in high density regions. We use this phenomenological
model for the variance of the halo bias relation.

Briefly, the mean of the bias relation from the MW model and our 
phenomenological modification of the \citet[][]{ShethL:99} 
formula for the variance\footnote{We only use the spherical model here
because a consistent implementation of the ellipsoidal model into the
phenomenological model for the variance is not straightforward.}, are given by
\begin{equation}
\langle N \rangle =\int{dm~ N(m,\delta_1 \mid M,\delta_0)} 
\label{eq:mean}
\end{equation}
and
\begin{equation}
\sigma^2= \langle N(N-1)\rangle + \langle N \rangle - \langle N \rangle^2,
\label{eq:variance}
\end{equation}
where:

\begin{eqnarray}
\langle N(N-1) \rangle &=& \int{dm_1\,dm_2\ N(m_1, \delta_1 \mid
M,\delta_0)} \nonumber\\ 
 & & ~~~N(m_2, \delta_1 \mid M-m_1,\delta') (1 +
A\overline{\xi}_2),
\label{eq:var}
\end{eqnarray}

\noindent $N(m,\delta_1 \mid M,\delta_0)$ denotes the average
number of haloes of mass $m$ identified at a given epoch $z_1$ [with a
critical overdensity for collapse $\delta_1 = \delta_c (1 + z_1)$] in
an uncollapsed spherical region of comoving volume $V$ with mass $M$ and
overdensity $\delta_0$, and $\delta'$ is the mass density contrast of
the fraction of the volume not occupied by the $m$ haloes. The
additional term $(1 + A \overline{\xi}_2)$ in the expression for the
variance accounts for the contribution from mass clustering  and has
been constructed as the simplest function of the variance of the mass
distribution with the property of having high values in overdense
regions and of being unity in homogeneous regions. 
As we will show below, a good fit to the simulation data
can be achieved by choosing ${\overline\xi}_2$ to be the 
second order moment of the mass distribution on the scale 
in consideration. In this case, we can write the term
$A{\overline\xi}_2=A{\overline\xi}_m(z_1)
\approx A D^2(z_1){\overline\xi}_m(0)$, where 
$D(z)$ is the linear growth factor normalized to one
at $z=0$. The constant $A$ is to be calibrated by simulations.

\section {Test by $N$-Body Simulations}
\label{sec:testing}

\subsection{Numerical Data}
\label{sec:data}

For this study we use the spatial distribution of dark matter
particles as well as of dark haloes from the \lcdm version
of the high resolution GIF N-body simulations \citep[for details
see][]{Kauffmann:99}. These simulations have $256^3$ particles in a
grid of $512^3$ cells, with a gravitational softening length of 
$20\kpch$. In the \lcdm case, the simulation assumes 
$\Omega= 0.3$, $\Omega_{\Lambda}=0.7$ and $h=0.7$. The initial power spectrum 
has a shape parameter $\Gamma = 0.21$ and is normalized 
so that the rms of the linear mass density in a sphere of 
radius $8\mpch$ is $\sigma_8=0.9$. The simulation box has a
side length $L=141\mpch$, and the mass of each particle
is $M_p=1.4\times 10^{10}~h^{-1}{\rm M_\odot}$. 

The halo catalogues have been created by the GIF project
\citep[][]{Kauffmann:99} using a friends-of-friends group-finder
algorithm to locate virialized clumps of dark matter particles in the
simulations outputs. They used a linking length of 0.2 times the mean
interparticle separation and the minimum allowed mass of a halo is 10
particles. In what follows, the mass of a halo is represented by the
number of particles it contains. 

We also use the galaxy catalogues constructed from the same 
simulations. The catalogues are limited to model galaxies with 
stellar masses greater than $\sim 2 \times 10^{10}~ h^{-1}\msun$. 
For further details about these catalogues and the galaxy formation 
models used in their construction see \citet[][]{Kauffmann:99}. 

The halo conditional probability $P_V(N_h \mid \delta_m)$ has been 
estimated for various samples and a number of 
sampling volumes $V$. For the presentation, we only use
samples of haloes selected at redshifts $z=3$, 1 and 0. 
Two kinds of analyses are performed. In the first case, 
halo counts-in-cells are estimated at the same time 
as when the haloes are identified, while in the second case
counts-in-cells are estimated at a given time for the central 
particles of the haloes identified at an earlier 
time. As a convention, we use $z_1$ to denote the redshift
at which haloes are identified, while using $z_0$ to denote
the redshift at which the counts in cells are estimated.
For example, a case with $z_0=0$ and $z_1=3$ means that haloes
are identified at redshift $3$ while the counts-in-cells are
estimated for their central particles at redshift $0$. 
The computations of the counts in cells are performed for
volumes of cubical cells with side lengths
1/32, 1/16, 1/8, and 1/4 times the side length of the
simulation box, corresponding to $4.4$, 8.8, 17.6 and 
$35.2\mpch$ in comoving units. The algorithm proposed by 
\citet[][]{Szapudi:99}, which allows an accurate determination 
of the probability function in a relatively short time, 
is applied to estimate the counts-in-cells on a  grid of 
$256^3$ cells. The conditional probability of finding $N_h$ 
haloes in a cell of volume $V$ given that the local mean mass 
overdensity has a value between $\delta_m$ and $\delta_m 
+\Delta\delta_m$ is computed from the counts-in-cells through 
\begin{equation}
P_V(N_h \vert\delta_m) = \frac{P(N_h,\delta_m)\,\Delta\delta_m}
{P(\delta_m)\Delta\delta_m},
\end{equation}
where $P_V(N_h,\delta_m)$ is the joint probability for finding $N_h$
haloes and a mass overdensity between $\delta_m$ and 
$\delta_m + \Delta\delta_m$ in a cell of volume $V$, 
and $P_V(\delta_m)$ is the distribution function for
the underlying mass density field. 

\subsection{The Conditional Probability}

\begin{figure*}
\includegraphics[width=\textwidth,height=\textwidth]{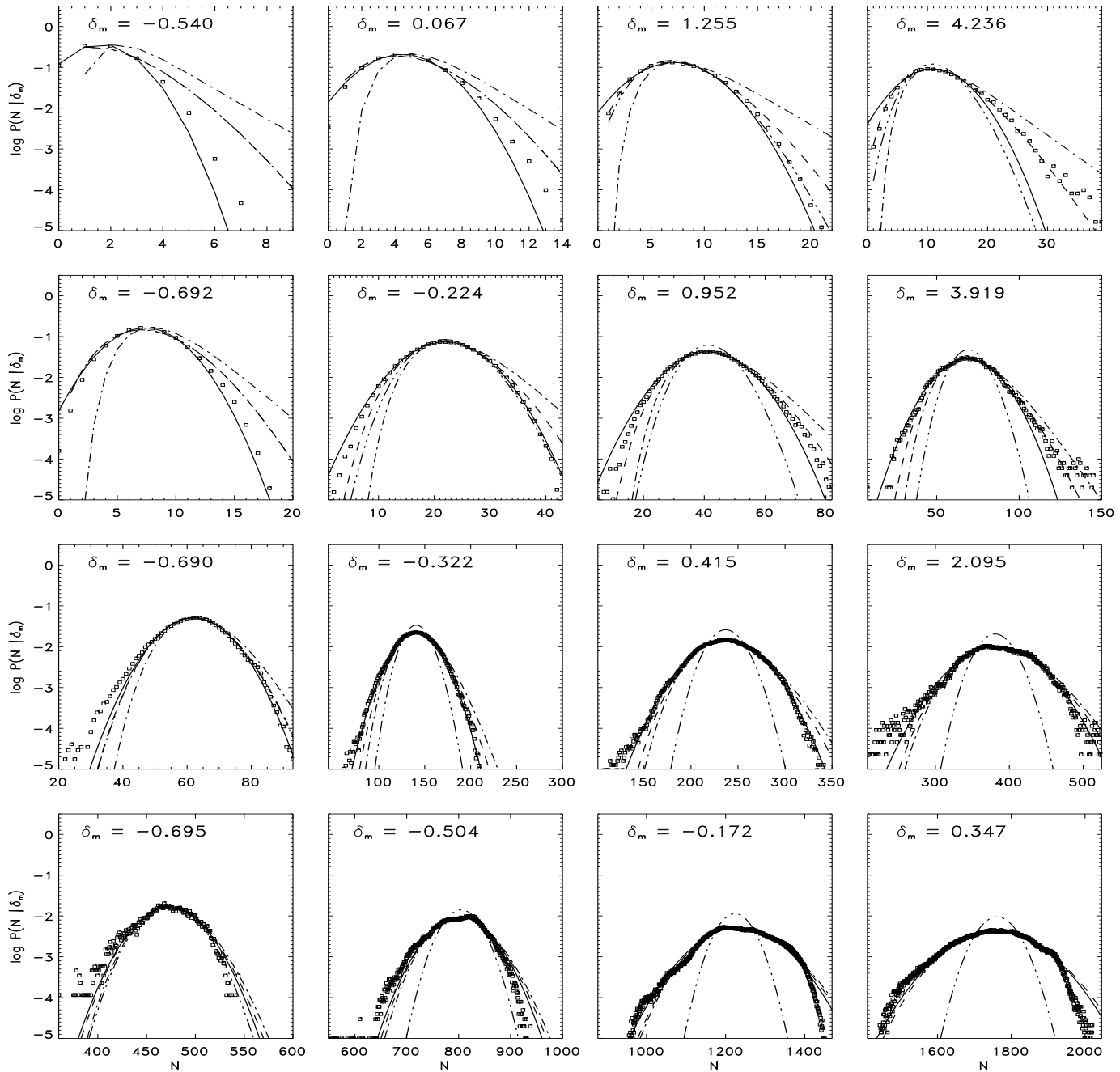}
\caption{Comparison between the conditional probability measured from
the simulations (squares) for present epoch haloes with masses greater
than 10 particles and the corresponding best fits of the Poisson
(dash-dot-dot-dot line), Thermodynamical (dashed line), Lognormal
(dash-dot line) and Gaussian (solid line) distribution functions. The
rows correspond, from top to bottom, to the sampling scales
$\ell = 4.4,~8.8,~17.6,~35.2~Mpc/h$, respectively. For each sampling scale
there are four plots corresponding to the local mass overdensity as
indicated in the labels.}
\label{fig:fit}
\end{figure*}

Figure \ref{fig:fit} shows the conditional probability  
$P_V(N_h\vert\delta_m)$ obtained from the simulations at several
representative values of $\delta_m$. The halo sample shown contains
all present-day haloes with masses greater than 10 particles. The
numerical conditional probabilities are compared with the
fits to Gaussian, Lognormal, Poisson and Thermodynamical
functions. From the figure it can bee seen that the Poisson
model is in general a poor description of the present time conditional
probability measured from the simulations, and that the Gaussian model
is overall a quite good assumption. The Lognormal and Thermodynamical
functions are a sort of intermediate functions, i.e they are not as
poor descriptors for the conditional probability as the Poisson
function is, but, on the other hand, they do not describe the features
of the conditional function as good as the Gaussian function does.
This result is valid for all the halo mass ranges under analysis and
all the scales tested. 

As discussed by \citet[][]{ColesF:91}, the widely used Poisson model
introduced by \citet[][]{Layzer:56} is based on the assumption that the
probability to find a given number of galaxies (or haloes) in a volume
is determined only by the local mass density field and that the
probability to find a galaxy (or halo) in an infinitesimal volume
$\delta V$ is proportional to the density at this volume element and
is independent of the probability to find another galaxy (or halo) in
a neighboring infinitesimal volume. Therefore, our finding that the
conditional probability for haloes has non-Poisson form suggests that
the deviation from Poisson of the halo-biasing process is due to
effects related to the probability for a halo to have neighbors, such
as volume exclusion, clustering and large-scale environments. Notice
that some of these effects are already included in the model for the
variance of the bias relation (section \ref{sec:model}).

\subsection{The Mean and Variance of Halo Bias}

Given that the Gaussian model is a reasonable fit to 
the conditional probability function, we now concentrate
on the mean and variance of this function, which
are the two quantities needed to specify a Gaussian 
distribution. In order to show deviations from the Poisson 
distribution, we consider the mean and the ratio between the 
variance and the mean:
\begin{equation}
1+\delta_h\equiv {\langle N\vert\delta_m\rangle\over {\overline n}V},
~~~~~~~
{{\rm variance}\over {\rm mean}}
\equiv {\sigma^2\over \langle N\vert\delta_m\rangle},
\end{equation}
where $\delta_h \equiv {{N_h}\over {{\overline n}V}} -1$ is the number
density contrast of haloes and ${\overline n}$ is their mean number
density.

\begin{figure*}
\centerline{\includegraphics[width=\textwidth]{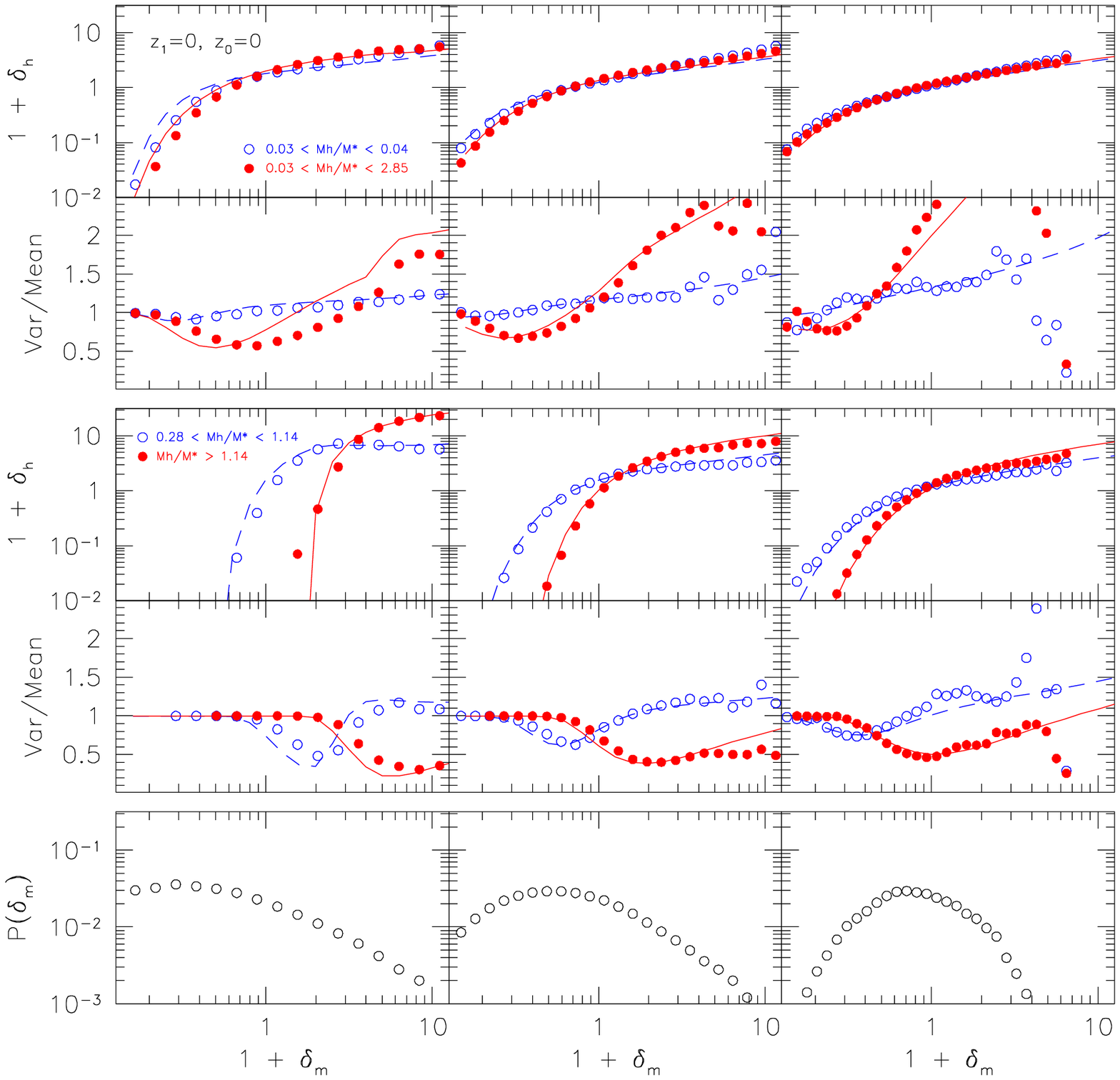}}
\caption{Theoretical predictions from the MW model for the
mean and from our proposed phenomenological modification of the
\protect{\citep[][]{ShethL:99}} model for the variance of the bias relation
(lines) compared with the corresponding quantities obtained from the
GIF \lcdm simulations (symbols). The columns correspond, from left to
right, to the cell sizes $l=4.4$, $8.8$ and $17.6~Mpc/h$. The two
upper panels show the mean of the bias relation (upper row in panel)
and the ratio between the variance and the mean of the bias relation
(lower row in panel) for the ranges of halo masses indicated in the
respective labels. The dashed and solid lines show the
theoretical predictions corresponding to the numerical data
represented by the open and filled circles, respectively. The mass
probability function at the respective scales is shown in the lowest
panel. The sample corresponds to haloes identified and analyzed at the
present epoch.} 
\label{fig:dhdm00}
\end{figure*}

\begin{figure*}
\centerline{\includegraphics[width=\textwidth]{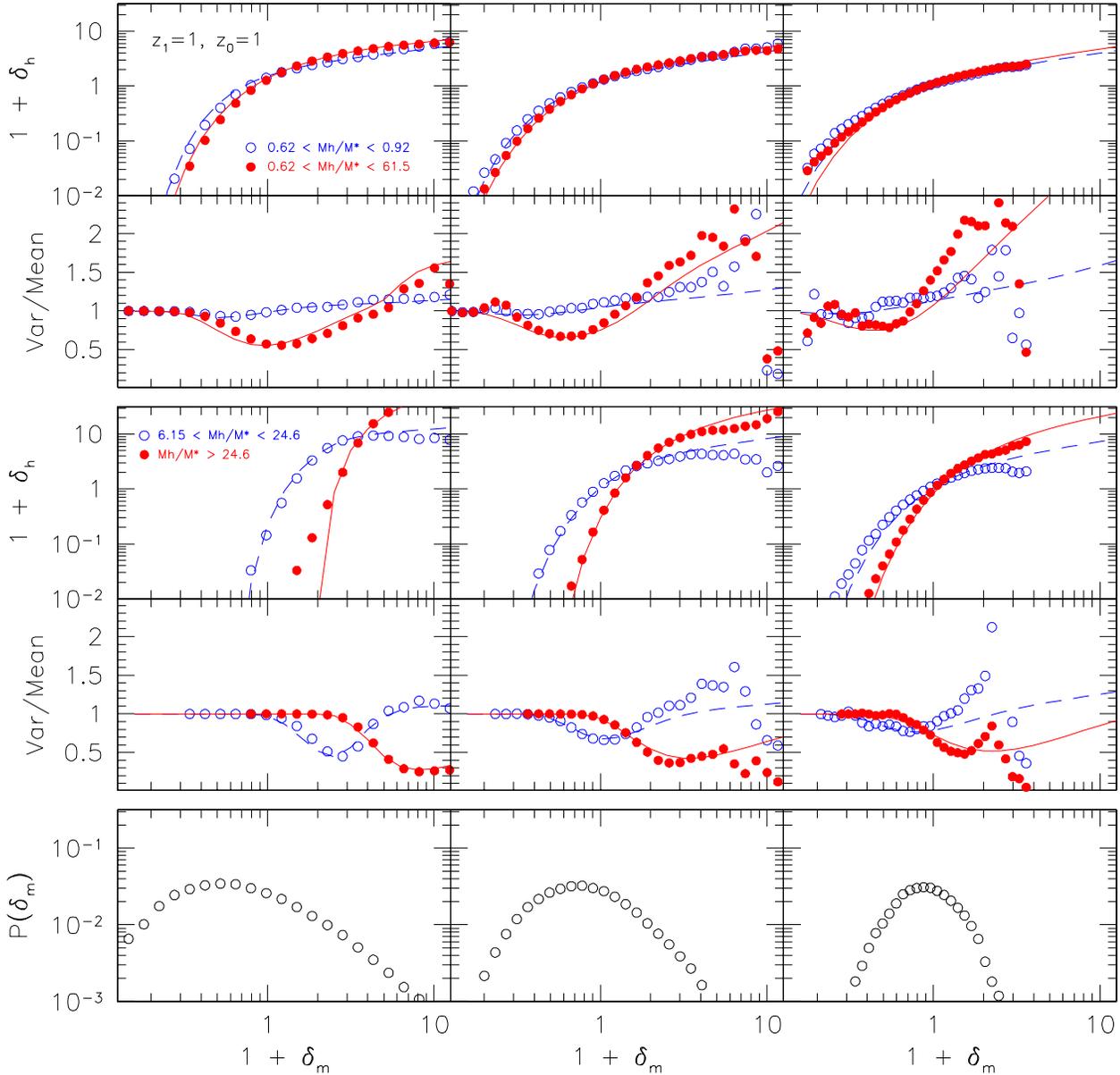}}
\caption{Same results as shown in figure (\ref{fig:dhdm00}) but for 
haloes identified and analyzed at redshift $z=1$.}
\label{fig:dhdm11}
\end{figure*}

\begin{figure*}
\centerline{\includegraphics[width=\textwidth]{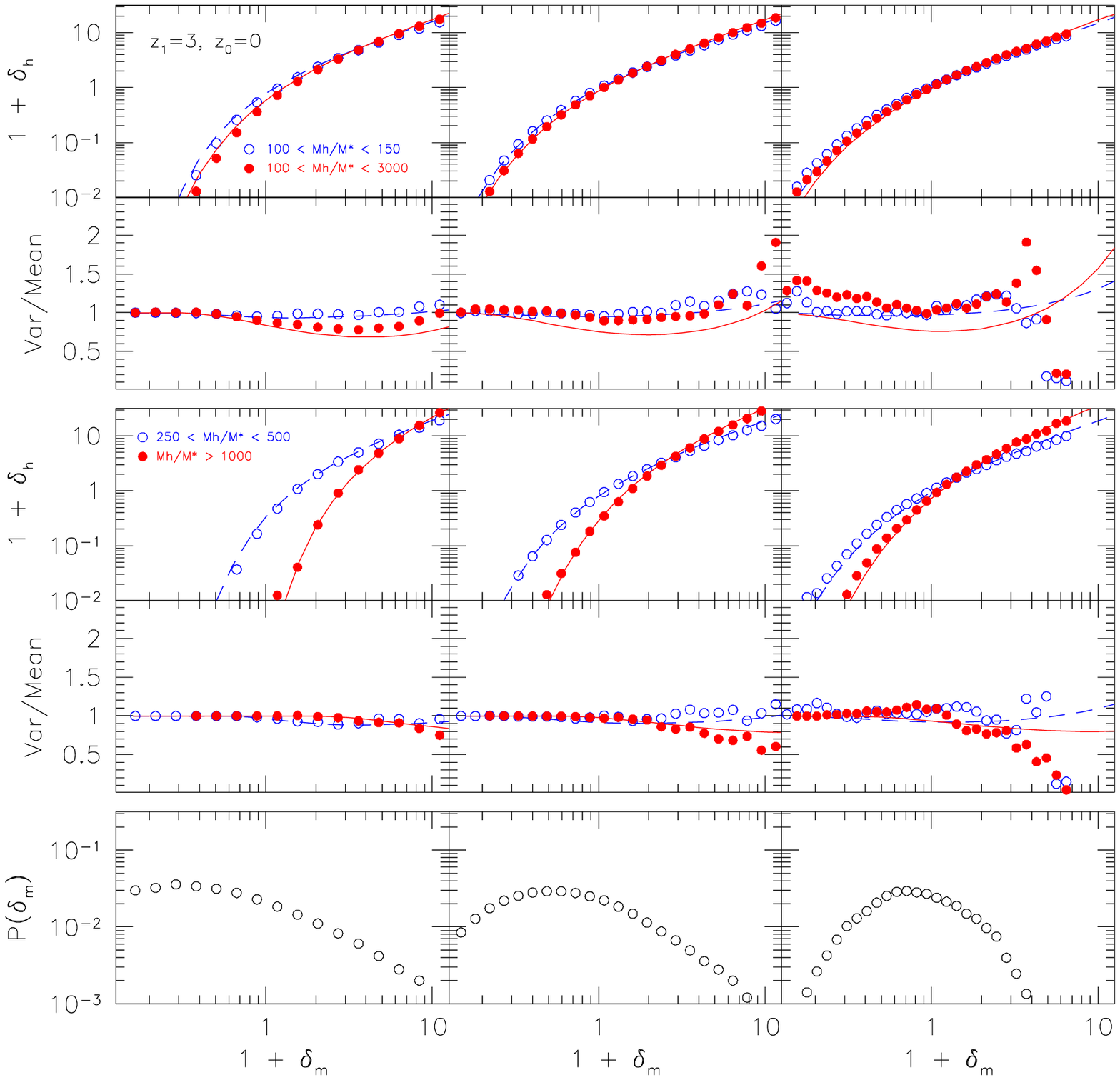}}
\caption{The same as in figure (\ref{fig:dhdm00}) but for the present
epoch descendants of haloes already formed at $z=3$.}
\label{fig:dhdm30}
\end{figure*}

Figures \ref{fig:dhdm00}-\ref{fig:dhdm30} show the results
given by the simulations. Results are shown for samples
in four representative mass ranges: a) a sample of low mass haloes, b)
a sample containing both low and high mass haloes, c) a sample of
intermediate mass haloes and d) a sample of high mass haloes. The
corresponding halo masses are shown in table 1.

\begin{table*}
\label{table:masses}
\centering
\begin{tabular}{l*{4}{c}}
\hline
Sample  & $z_1 = 0$  & $z_1 = 1$ & $z_1 = 3$  \\
\hline
a) & $M_h=20$--30 particles   & $M_h=20$--30 particles & $M_h=20$--30 particles \\
   & $M_h/M_{\star} = 0.03$--0.04  & $M_h/M_{\star}=0.62$--0.92 & $M_h/M_{\star}=100$--150 \\
\hline
b) & $M_h=20$--2000 particles & $M_h=20$--2000 particles &
$M_h=20$--600 particles \\
   & $M_h/M_{\star} = 0.03$--2.85 & $M_h/M_{\star}=0.62$--61.5 & $M_h/M_{\star}=100$--3000 \\
\hline
c) & $M_h=200$--800 particles & $M_h=200$--800 particles & $M_h=50$--100 \\
   & $M_h/M_{\star} = 0.28$--1.14 & $M_h/M_{\star}=6.15$--24.6 &
$M_h/M_{\star}=250$--500 \\
\hline
d) & $M_h > 800$ particles    & $M_h > 800$ particles   & $M_h > 200$ \\
   & $M_h/M_{\star} > 1.14$       & $M_h/M_{\star} > 24.6$ & $M_h/M_{\star} >
1000$ \\
\hline
\end{tabular}
\caption{Ranges of halo masses corresponding to the samples shown
in figures \ref{fig:dhdm00}-\ref{fig:dhdm30}. a) sample of low mass
haloes, b) sample containing both low and high mass haloes, c) sample
of intermediate mass haloes and d) sample of high mass
haloes. $M_{\star}$ is defined by $\sigma(M_{\star}) = 1.68$.} 
\end{table*}

One sees that the ratio variance/mean 
shows a Poisson-like behavior (i.e. $\sim 1$) for low values 
of $\delta_m$. This ratio becomes sub-Poisson (i.e. $< 1$) at
intermediate values of $\delta_m$, and super-Poisson ($>1$) 
for high values of $\delta_m$. The exact change of the variance/mean
ratio with $\delta_m$ depends on halo mass: 
the sub-Poisson variance extends to higher values of
$\delta_m$ for samples with higher halo masses. 
The volume-exclusion effect is reduced for the descendants of 
haloes identified at an earlier epoch and the variance/mean 
ratio approaches the Poisson value for the descendants of haloes 
selected at early times (see Figure \ref{fig:dhdm30}). 

The curves in Figures \ref{fig:dhdm00}-\ref{fig:dhdm30} 
show model predictions. The mean bias relations given by
the simulations are well described by the model of MW, 
confirming earlier results. The behavior of the 
variance/mean ratio is also reasonably well reproduced, 
when the constant $A$ in equation (\ref{eq:variance})
is chosen to be 0.05 (as given by the fit to the bias relation 
for present-day haloes in the simulation). Thus, sub-Poisson variance
can be caused by halo exclusion while the super-Poisson variance at
high $\delta_m$ may be explained by the clustering of mass at the time
of halo identification. The model for the variance begins to fail at
very high values of $\delta_m$. But since cells with such high
densities are only a tiny fraction of all cells, this failure is not
important.

\subsection{Bias Relation for Model Galaxies}

\begin{figure}
\centerline{\includegraphics[width=\columnwidth]{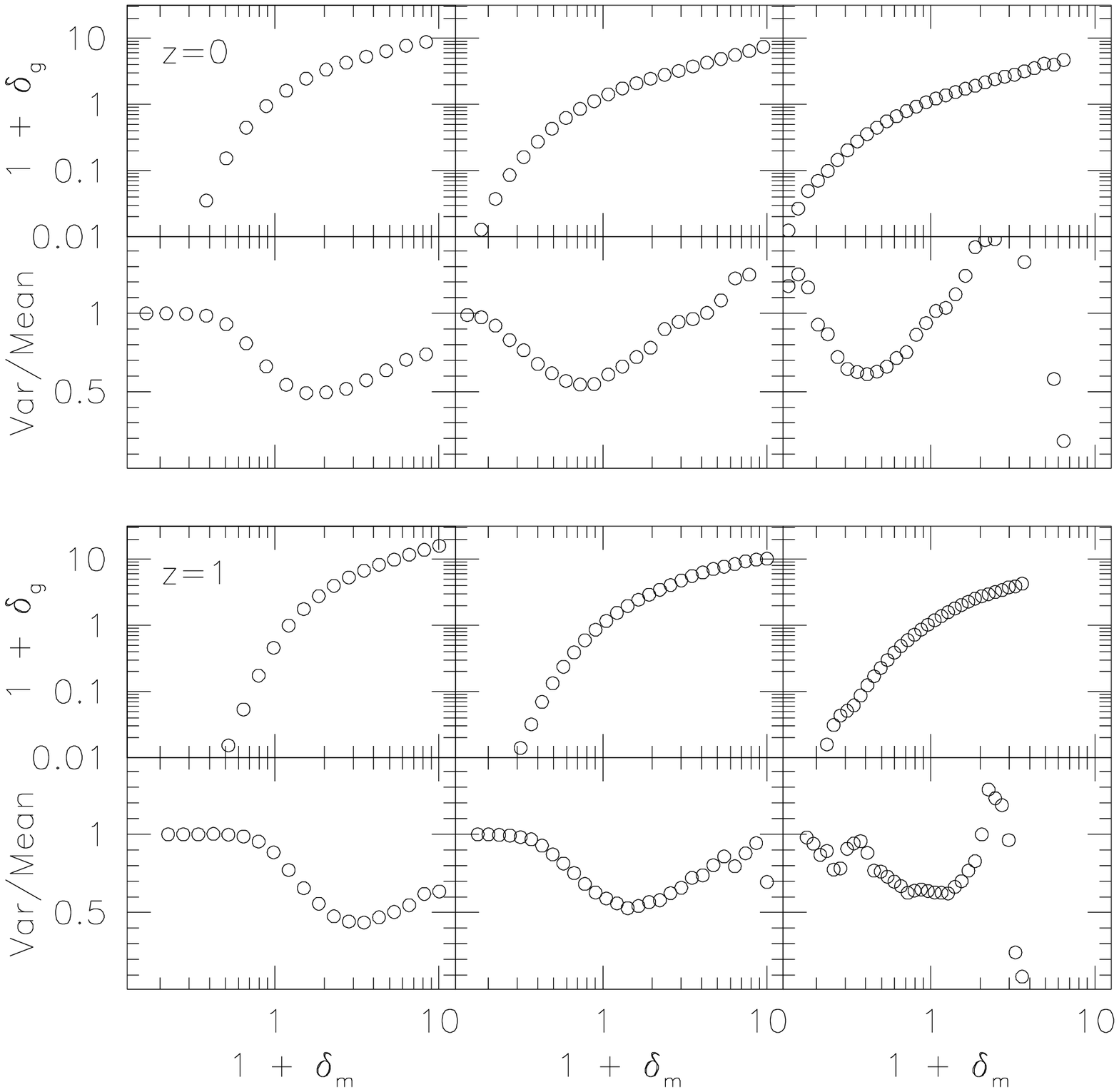}}
\caption{Mean bias relation and ratio between the variance and the
mean of the bias relation of galaxies obtained from the simulations
using semi-analytical models of galaxy formation. We show model
galaxies at the present epoch (upper panel) and at redshift $1$ (lower
panel). The mean and ratio between the variance and the mean of the
bias relation are shown in the top and bottom rows in each panel,
respectively. At each epoch the cubical cells of side 
length, from left to right, $l=4.4,~8.8,~17.6~h^{-1}~Mpc$ are shown.}
\label{fig:dhdm_gal}
\end{figure}

 We have also estimated the mean and variance of the bias relation 
between model galaxies, from the GIF simulations, with stellar
masses greater than $\sim 2\times 10^{10} h^{-1}\msun$ and the
underlying mass density, with the results shown in Figure
\ref{fig:dhdm_gal}. Interestingly, the variance/mean ratio in the
galaxy-mass bias relation also exhibits significant sub-Poissonian
behavior, implying that the effect of volume exclusion is also
important for the spatial distribution of galaxies. One possible
reason for this is that many of the galaxy-sized haloes may host only
one galaxy and the galaxy distribution inherits a considerable
fraction of the exclusion effects from the distribution of their host
haloes. 

If this is true for real galaxies, it has important
implications for the interpretations of galaxy clustering,
as we will see in Section \ref{sec:summary}.

\subsection{The Count-in-Cell Function of Dark Haloes}

\begin{figure}
\centerline{\includegraphics[width=\columnwidth]{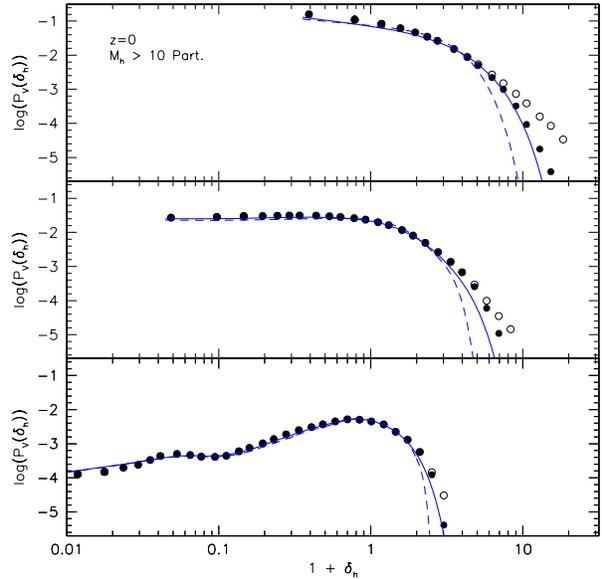}}
\caption{Halo count-in-cell functions for a sample of
present day haloes with masses greater than 10 particles. The circles
correspond to the probability function obtained from the simulations
and the lines to the semi-analytically reconstructed count-in-cell
function using spherical collapse approach. The solid and dashed lines
show the reconstructed functions using a Gaussian and a Poissonian
form for the conditional probability function, respectively. The open
circles show the simulated halo count-in-cell functions obtained from
the complete mass and conditional probability functions, while the
filled circles show the simulated mass count-in-cell functions
obtained from the mass and conditional probability functions truncated
at high values of the mass density contrast. The boxes correspond,
from top to bottom, to the scales $l=4.4,~8.8,~17.6~h^{-1}~Mpc$.}
\label{fig:fig_halo}
\end{figure}

As an additional test for the bias model, we use
the simulation result for $P_V(\delta_m)$ 
and the theoretical model for $P_V(N_h\vert\delta_m)$ 
[i.e. a Gaussian conditional probability function with the 
mean and variance given by equations (2)-(4)]
to reconstruct the counts-in-cells functions for haloes 
by using equation (\ref{eq:pdf}). As we only want to test the 
model of the bias relation, we do not use theoretical
models for $P_V(\delta_m)$, although such models do exist 
[e.g. the model of \citet[][]{Sheth:98} based on excursion set 
approach, and the Lognormal model used in
\citet[][]{ColesJ:91}]. Since the probability functions obtained from
the simulations are quite noisy at very high values of $\delta_m$ and
the model predictions in this regime may fail, we truncate our
computations at a given high value of $\delta_m$ [$\delta_m^{max} =
10$ at the scales $l=4.4\ {\rm and}\ l=8.8~h^{-1}~Mpc$, 
and  $\delta_m^{max} = 3$ at $l=17.6~h^{-1}~Mpc$], which correspond
to the low-probability tail of the mass probability function, as can
be seen clearly in the lower panel in figure (2). For
comparison we also reconstruct the halo count-in-cell functions using 
a Poissonian form for the conditional function, with the mean given by
equation (2). In Figure (\ref{fig:fig_halo}) we compare the reconstructed
halo count-in-cell functions for present-day haloes containing more
than 10 particles with the corresponding functions obtained directly
from the simulations. Clearly, the model matches the simulation
results remarkably well. The halo count-in-cell functions
reconstructed using a Poissonian form for the conditional probability
function depart from the corresponding numerical values in the
low-probability, high density tail.

The halo count-in-cell functions obtained through this 
approach can be used to calculate the high-order moments, 
such as skewness and kurtosis, of halo distributions. 
This application will be presented in detail 
in a forthcoming paper.

\section {Discussion and Summary}
\label{sec:summary}   

In this paper, we have analyzed in detail the conditional probability
function $P_V(N\vert\delta_m)$ to understand the stochastic nature of
halo bias. We have found that in high-resolution $N$-body simulations 
this function is well represented by a Gaussian model, and that a
Poisson model is generally a poor approximation. That means that the
galaxy biasing process, as well as the halo biasing process, is not
only determined by the local value of the mass density field, but also 
by other local quantities, such as clumpiness, and by non-local
properties, such as large-scale tidal field.

We have shown that a simple, phenomenological model can be 
constructed for $P_V(N\vert\delta_m)$. This allows one to 
construct a theoretical model for the full count-in-cell
function for dark haloes. The galaxy distribution in the cosmic 
density field predicted by semi-analytic models of galaxy 
formation shows similar stochastic behavior to that of the haloes,
implying that galaxy distribution is not a Poisson
sampling of the underlying density field.

These results have important implications in the interpretations
of galaxy clustering in terms of the underlying density 
field. For example, the quantity conventionally used to 
characterize the second moment of counts-in-cells is
defined (here for dark halo) as
\begin{equation}
\kappa_2 (R)
= {\langle (N-{\overline n}V)^2\rangle\over ({\overline n}V)^2}
-{1\over ({\overline n}V)},
\end{equation}
where the second term on the right-hand side is
to subtract Poisson shot noise \citep[e.g.][]{Peebles:80}.
With the use of equation (\ref{eq:pdf}), it is easy 
to show that
\begin{eqnarray}
\kappa_2(R)
&=& {1\over ({\overline n}V)^2}
\int\langle N\vert\delta_m\rangle^2 P_V(\delta_m)\,d\delta_m \nonumber\\
&+& {1\over ({\overline n}V)^2}
\int\left[\sigma^2-\langle N\vert\delta_m\rangle\right]
P_V(\delta_m)\,d\delta_m-1\,.
\end{eqnarray}
Thus, even if haloes trace mass on average, i.e.
$\langle N\vert\delta_m\rangle\propto \delta_m$, 
this quantity is not equal to the second moment for the mass, 
because the second term on the right-hand side is 
generally non-zero. Furthermore, the non-Poissonian behavior of
the bias relation might imply that the (Poisson) shot-noise
corrections usually applied at estimating higher-order moments of the
galaxy distribution are not completely correct and therefore
interpretations of skewness and kurtosis might change considerably, at
least at the scales where shot-noise terms are not too small. This
issue needs to be investigated in more detail. Thus, in order to infer
the properties of the mass distribution in the Universe from
statistical measures of the galaxy distribution, it is necessary to
understand the stochastic nature of galaxy biasing.

As discussed in \citet[][]{DekelL:99}, the stochasticity in galaxy
biasing not only affects the interpretation of the moments of the
galaxy distribution, but also affects the interpretation of other
statistical measures of galaxy clustering, such as redshift
distortions, the cosmic virial theorem and the cosmic energy
equation. With the results shown in the present paper, one can
model quantitatively many of these effects. 
       
\section*{Acknowledgments}

We are grateful to Guinevere Kauffmann for a careful reading of the
manuscript.  We thank the GIF group and the VIRGO consortium for the
public release of their N-body simulation data 
({\tt www.mpa-garching.mpg.de/Virgo/data\_download.html}).   
R. Casas-Miranda acknowledges financial support from the ``Francisco
Jos\'e de Caldas Institute for the Development of Science and
Technology (COLCIENCIAS)'' under its scholarships program. RKS is
supported by the DOE and NASA grant NAG 5-7092 at Fermilab. He also
thanks the Max-Planck Institut fuer Astrophysik for hospitality at the
initial stages of this work. We also thank our referee, Peter Coles,
for useful comments and suggestions. 

\bibliographystyle{mn2e}
\bibliography{Mycitations}
\clearpage

\end{document}